\documentclass[letterpaper, 10 pt, conference]{ieeeconf}  % Comment this line out if you need a4paper

\IEEEoverridecommandlockouts                              % This command is only needed if 
\usepackage{graphics} % for pdf, bitmapped graphics files
\usepackage{epsfig} % for postscript graphics files
\usepackage{mathptmx} % assumes new font selection scheme installed
\usepackage{times} % assumes new font selection scheme installed
\usepackage{amsmath} % assumes amsmath package installed
\usepackage{amssymb}  % assumes amsmath package installed
\usepackage[english]{babel}
\usepackage{hyperref}
\usepackage{dblfloatfix} 
\usepackage{cite}

\usepackage{algorithm,algpseudocode}
\makeatletter
\newcommand{\algmargin}{\the\ALG@thistlm}
\makeatother
\newlength{\whilewidth}
\algdef{SE}[parWHILE]{parWhile}{EndparWhile}[1]
  {\parbox[t]{\dimexpr\linewidth-\algmargin}{%
     \hangindent\whilewidth\strut\algorithmicwhile\ #1\ \algorithmicdo\strut}}{\algorithmicend\ \algorithmicwhile}%
\algnewcommand{\parState}[1]{\State%
  \parbox[t]{\dimexpr\linewidth-\algmargin}{\strut #1\strut}}

\usepackage[draft]{pgf}
\usepackage{shapepar}
\usepackage{tikz}
\usetikzlibrary{shapes.geometric}

\usepackage{verbatim}  % defines comment environment - JH

\title{\LARGE \bf
Maneuver Regulation for Accelerating Bodies in Atmospheric Environments 
}

\author{Juan-Pablo Afman$^{1}$, Eric Feron$^{1}$, John Hauser$^{3}$ % <-this % stops a space
\thanks{$^{1}$J.P. Afman, E. Feron are with the Department of Aerospace Engineering at the Georgia Institute of Technology, Atlanta, GA 30332, USA.}
\thanks{$^{3}$John Hauser is a private consultant.}% {\tt xx@xx.edu}}
}

\begin{document}

\maketitle
\thispagestyle{empty}
\pagestyle{empty}

%%%%%%%%%%%%%%%%%%%%%%%%%%%%%%%%%%%%%%%%%%%%%%%%%%%%%%%%%%%%%%%%%%%%%%%%%%%%%%%%
\begin{abstract}
In order to address the need for an affordable reduced gravity test platform, this work focuses on the analysis and implementation of atmospheric acceleration tracking with an autonomous aerial vehicle. As proof of concept, the vehicle is designed with the objective of flying accurate reduced-gravity parabolas. Suggestions from both academia and industry were taken into account, as well as requirements imposed by a regulatory agency. The novelty of this work is the Proportional Integral Ramp Quadratic PIRQ controller, which is employed to counteract the aerodynamic forces impeding the vehicles constant acceleration during the maneuver. The stability of the free-fall maneuver under this controller is studied in detail via the formation of the transverse dynamics and the application of the circle criterion.  The implementation of such a controller is then outlined, and the PIRQ controller is validated through a flight test, where the vehicle successfully tracks Martian gravity 0.378 G's with a standard deviation of 0.0426.  %The vehicle employs state transition logic to perform the maneuver fully autonomously, from takeoff to landing, while monitoring for mission and hardware faults at each time step. 
\end{abstract}

%%%%%%%%%%%%%%%%%%%%%%%%%%%%%%%%%%%%%%%%%%%%%%%%%%%%%%%%%%%%%%%%%%%%%%%%%%%%%%%%

\section*{Motivation}
Enabling safe and reliable reduced-gravity environments for short temporal intervals is the objective behind this work. A wide range of scientific disciplines employ reduced-gravity conditions as a tool to investigate phenomena of processes in the absence of Earth's gravitational field. These disciplines include both life sciences as well as physical sciences. When gravity is removed, other forces such as surface tension and capillary forces, become predominant and drive a different system dynamics. Gravity is a physical parameter, that together with pressure and temperature, define the state of a system. Historically, major breakthrough and innovations were achieved when systems were studied, for example, at low temperatures. The duration of the reduced gravity periods can range from a matter of seconds to months, depending on the application and platform employed~\cite{Prasad}. The scope of this work targets the development of an Earth-based vehicle with mission critical safety and autonomy,  and can deliver adequate gravitational fields to the hands of the students and researchers. Although reduced-gravity research has proven itself useful for public benefit through breakthroughs in the pharmaceutic, metallurgy, communications, electronics~\cite{cancer,insulin,metal,comm,semi,mems}, the means of providing an affordable and reliable unmanned test facility remains an unsolved problem. 

Two common Earth-based solutions employed for research experiments include drop towers \cite{nature1,Bremen,Glenn} and the the parabolic manned-aircraft\cite{vomit,ESA}. While drop towers can offer anywhere from 1.4 and up to 9 seconds of microgravity, the few towers capable of providing reduced-gravity, like the Bremen tower \cite{Bremen}, do so through centrifuges, which 1) are still limited to the drop time of a microgravity experiment and 2) induce a gravitational gradient across the experiment. Conceptual efforts to overcome this limitation have been proposed in recent work through the use of magnetism, although they have not been implemented in practice to this date~\cite{concept}. Furthermore, a single 5.18 second drop in the Zero-Gravity Research Facility at NASA Glenn, can cost up to \$9,000 U.S. dollars, and researchers are limited to two drops per day~\cite{Glenn}. Although there are benefits to the drop tower infrastructure, the major drawback is the large upfront cost for the infrastructure, its maintenance, operator requirements, and capacity; if a large number of users must be accommodated at the same time, a facility's accessibility diminishes considerably.

Perhaps the most common way of performing Earth-based reduced gravity experiments is through parabolic flight, which is is capable of providing an entire spectrum of gravitational environments for approximately 17-23 sec. intervals. Due to the special parabolic flight aircraft instrumentation, crew, and maintenance required, the operational costs associated with this platform are inhibiting for most researchers. The Zero Gravity Coorporation\textsuperscript{\textregistered} is currently the only commercial provider of microgravity flight services in the United States. According to the Zero-G Weightless Lab\textsuperscript{\textregistered} pricing information, single flight costs range from \$7,150  U.S. dollars + 5\% tax per flight for a hand held object, to \$38,500 + 5\% tax per flight for a larger experiment with a footprint no larger than 10 $ft^2$\cite{ZeroG}. The other obvious draw back expressed regular users is the lack of repetition, as experiments often fail in flight and experiment data is not aquired. Even at these high costs, NASA Inspector General Paul K. Martin reported that the Zero Gravity Coorporation has been incapable of delivering appropriate reduced-gravity environments~\cite{audit}. The Zero Gravity Corporation\textsuperscript{\textregistered} later pointed out that it had made significant investments in configuring its plane to NASA specifications and training pilots to conduct reduced gravity flights, which at minimum highlights the importance of feedback control for performing parabolic maneuver.

Although limited in scope, many studies can be performed with only a few seconds of reduced-gravity. In order to fulfill the gap faced by researchers from many fields, the development of a multirotor-based platform for reduced-gravity environments with increased availability, reliability, and reduced costs, is formalized. 

\section*{Concept Review and Requirement Definition}

\subsection*{Review of prior art}
There are major benefits, both financial and logistics, for enabling reduced gravity environments by means of removing the pilot from the vehicle, and decreasing infrastructure and operational costs. Some of these benefits include mission reliability, quick turn-around time, and minimum maintenance. These advantages have been the major incentive behind the work of several pioners, who have dedicated their efforts towards enabling autonomous reduced gravity environments over the past 22 years \cite{Ballistocraft,Freewing,jet,Kdai,JSC,zrhog}. 

To the best knowledge of the authors, every unmanned attempt to this date has been fixed a fixed winged aircraft, and sadly each attempt has failed to produce reliable and repeatable reduced-gravity environments. It is noted here that these failures are not due to the vehicle configuration, but rather that they have fallen victims of attempting to perform gravitational tracking in atmospheric environments through the means of a single Proportional Integral control law. From classical control theory, it is well understood that a single integrator is incapable of dynamically inverting the quadratic growing disturbance; a disturbance that rises due to the non linearity associated with aerodynamic forces while performing the maneuver.

In 1995, Meshland~\cite{Ballistocraft} published his experimental work on the BAR-1 aircraft, and the conceptual design of the BAR next. Cancellation of lift was obtained by controlling the elevator. The BAR-1 employed a feedback control law designed to minimize the error between a desired G set-point, and the actual measurement. Furthermore, this control input was augmented by a feed forward controller that too into account the G set-point, and the vehicle's forward airspeed. Unfortunately, the experimental results for both micro-gravity and reduced-gravity illustrate the controller's inability to reliably compensate to produced acceptable experimental conditions. No further work was found regarding the BAR next concept. In 2006, Kraeger~\cite{Freewing} proposed augmenting the BAR next concept previously conceptualized by Meshland, through the conceptual design of a free-wing unmanned aircraft. In his work, he shows that for this configuration, a relatively simply PI control architecture is sufficient to fly highly accurate reduced-gravity trajectories by applying torque to the wing. Unfortunately, this work never left the conceptual design phase, and hence validation of his claims were never confirmed experimentally. In 2009, Hofmeister and Blum~\cite{jet} proposed a jet-powered reduced-gravity test bed employing a modified unmanned aircraft from UR-Modellbau. The G-regulating feedback control structure was a simple PID, where the integral component was chosen to regulate the continuous set point deviation. Unfortunately, the experimental results for a micro-gravity maneuver show a 0.4 G RMS, which they claim is due to the aircraft being incapable of adjusting its absolute airspeed due to the vehicle's inertia. In 2010, Higashino and Kozai~\cite{Kdai} attempted the single integrator feedback control law that had failed in previous work, only to find similar experimental results. In 2011, a project led by Jeffery Fox from NASA Johnson Space Center~\cite{JSC} effort proposed a reduced-gravity platform employing a commercially-available Troybuilt “DV8R” remote-controlled jet aircraft. To the best of the author's knowledge, feedback control was never implemented and the work was never made publicly availabe. In 2016, Hathaway also proposed the use of a PID feedback control architecture, but his work never left the simulation phase.\cite{zrhog}

\subsection*{Requirements and Configuration}
A product discovery effort led by the research team, composed of interviews performed at over a dozen universities, research institutions and private organizations, revealed the requirements and constraints involved in the the development of an autonomous reduced-gravity platform. These are now itemized and described below. 

\begin{itemize}
  \item Affordability: Vehicle must cost less than \$25,000 USD.
  \item Reliability: Vehicle must be capable of producing repeatable and reliable results. 
  \item Safety and Convenience: Vehicle must address inconveniences faced by other platforms while providing the same level of safety.
  \item Time Exposure: Vehicle must be able to provide at minimum $2$ sec. of reduced-gravity conditions.
  \item Payload: Vehicle must be able to carry a 0.5 $kg$ payload and a volume of at least 10 $cm^3$
  \item Regulatory Constraints: Vehicle must not violate regulations relative to the operation of civilian drones.
\end{itemize}

In order to further define the reliability factor, flight specifications employed by NASA and the Zero Gravity Corporation\textsuperscript{\textregistered} were employed. Table 1 shows the flight specifications and requirements employed by NASA for judging the adequacy of reduced gravity time offered by the Zero Gravity Corporation\textsuperscript{\textregistered}.
\begin{table}[h]
\begin{center}
    \caption{Flight Specification Requirements \cite{audit}}
    \begin{tabular}{c c}
    \hline \linebreak[2]
    Reduced-Gravity (g's) & Continuous Time (sec) \\
    \hline %\linebreak[1]
0.00 g +/- 0.02  & 10\\
	0.00 g +/- 0.05  & 17 \\
	0.10 g +/- 0.05  & 20\\
	0.16 g +/- 0.05  & 20 \\
	0.20 g +/- 0.05  & 20\\
	0.30 g +/- 0.05  & 20 \\
	0.38 g +/- 0.05  & 20 \\
	0.40 g +/- 0.05  & 20\\
	0.50 g +/- 0.05  & 20\\
    \hline
    \end{tabular}
    \label{ZGTable}
\end{center}
\end{table}

The decision to use an autonomous UAV was made since it fits perfectly within the requirements and constraints discovered during the product discovery process. From a financial standpoint, most UAVs are mechanically simple and massively available on the commercial market free of restrictions, which makes them available and affordable to any academic research institution for open research, which makes them possible to meet the \$25,000 USD price tag. Second, autonomy and feedback control can be implemented in order to produce repeatable and reliable results on every flight. With fault detection algorithms executing at 333 $Hz$, it is possible to address the adequate level of safety requested. Furthermore, placing these vehicles on the hands of students and researchers covers the inconveniences face with all other Earth-based platforms. Lastly, time exposures, payloads, and regulations can all be met through proper vehicle sizing. Federal regulations posed by the US Federal Aviation Administration permits its use as long as it remains within eyesight, under 400 $ft$ (approximately 120 $m$) and with a mass lower than 55 $lbs$ (approximately 25 $kg$). In order to increase the convenience factor, the choice for an unmanned vehicle capable of vertical takeoff-and-landing VTOL implies minimum real-state requirements.  %In Germany, the maximum allowable altitude is 500 $ft$ and maximum mass is 25 $kg$.

\begin{figure}[h]
\begin{center}
\includegraphics[width=\columnwidth]{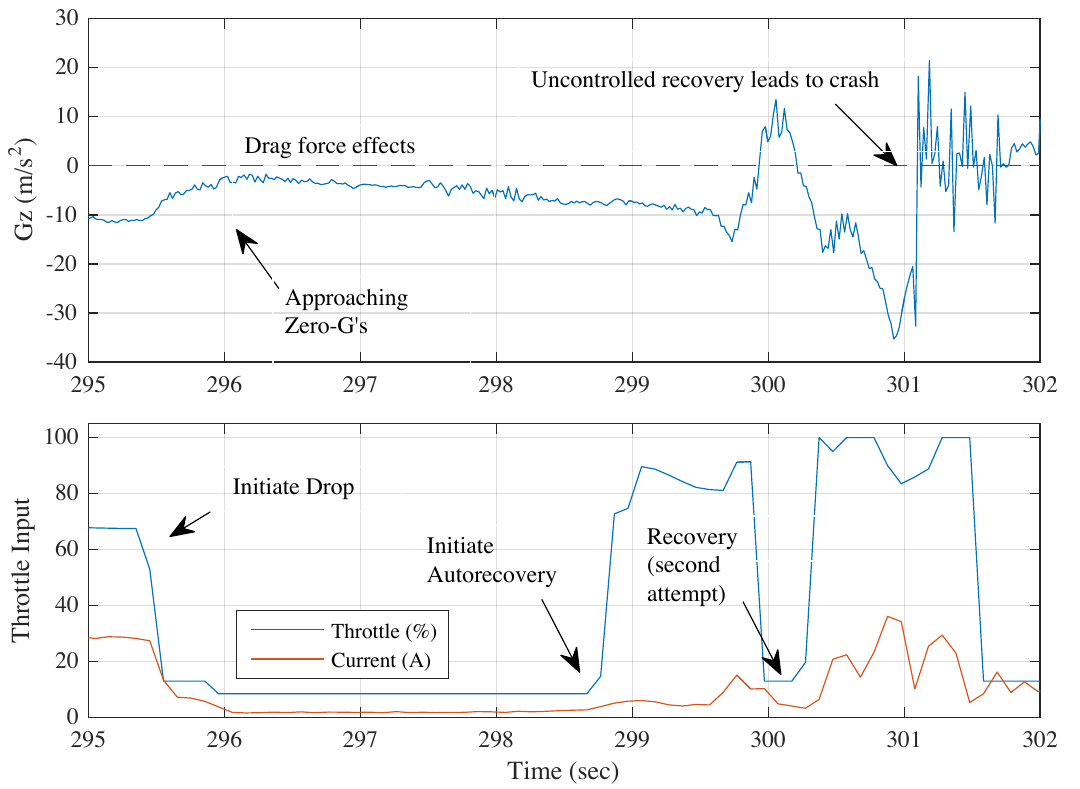}
\caption{Zero G tracking attempt with fixed pitch quadrotor}
\label{fp}
\end{center}
\end{figure}

Several modifications to the conventional multi-rotor architecture are necessary to obtain a vehicle capable of delivering proper reduced-gravity environments. Tracking a parabolic trajectory requires the use of both positive and negative thrust forces in order to compensate aerodynamic drag in ascent and descent portions of the parabolic flight; something that a conventional fixed-pitch multi-rotor simply cannot perform. Furthermore, since the thrust profile on a fixed-pitch quadrotor is a quadratic function of the propeller rotational velocity, control authority tends to zero as the thrust required goes to zero and hence, vehicle instability occurs. Fig. \ref{fp} shows experimental data for a fixed-pitch multi-rotor attempting to track a reduced-gravity profile. 

It is noted that both the slow responsiveness of fixed-pitch rotors, due to rotational inertia, as well as aerodynamic drag forces, prevent the vehicle from properly tracking a constant acceleration field. Furthermore, with thrust forces nearly zero, the the fixed-pitch vehicle is incapable of regulating its attitude during the maneuver and hence the flight test results end up with the loss of a vehicle.

\begin{figure}[h]
\begin{center}
\includegraphics[width=6cm]{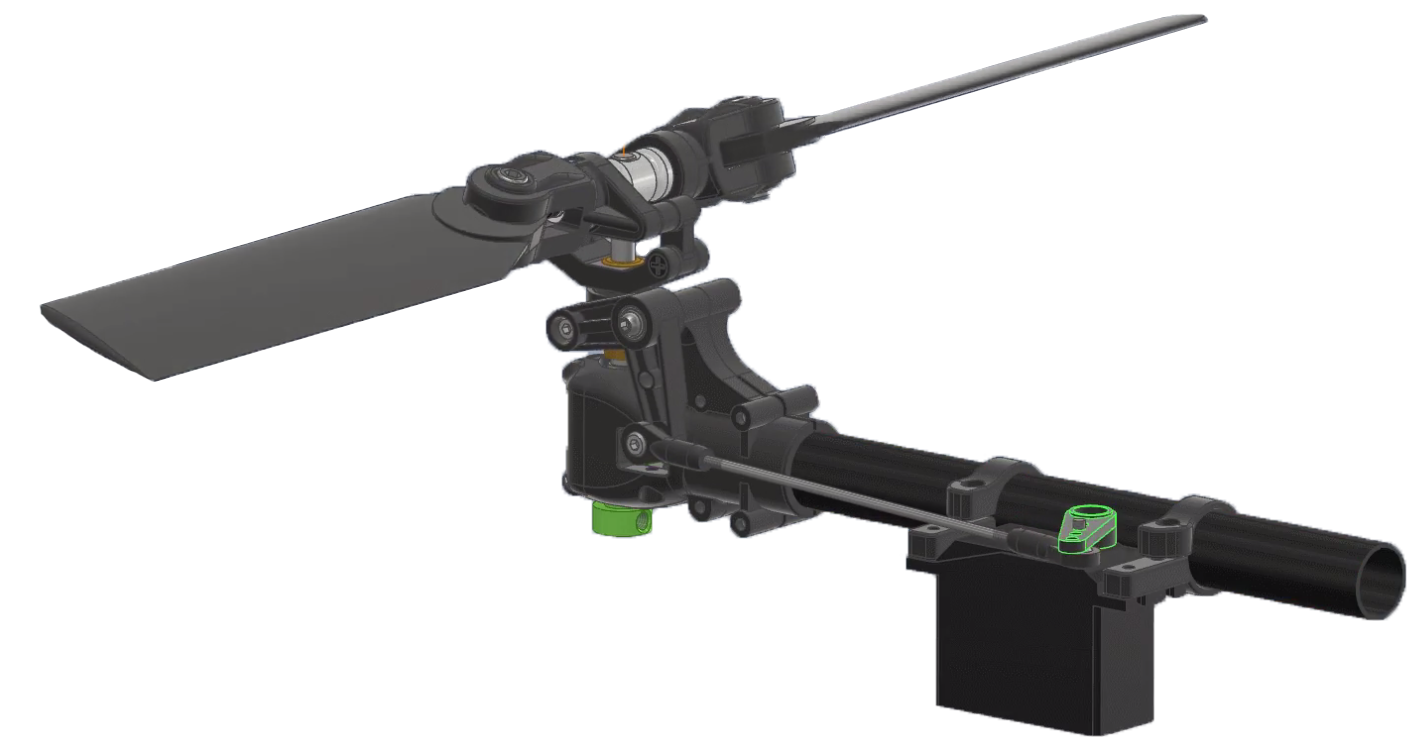}
\caption{Variable pitch mechanism}
\label{vp}
\end{center}
\end{figure}

To overcome the issues observed during fixed-pitch multi-rotor experiments, the research team looked towards the benefits of variable-pitch multi-rotors. This modification in the approach, which employs a constant rotor speed and $n$ independent pitch actuators, results in a responsive system with approximately 5 times control bandwidth, which is capable of fighting drag disturbances independent of their direction through positive and negative thrust deflections, all while maintaining attitude control authority independent of the thrust required during a reduced-gravity tracking flight. Figure~\ref{vp} illustrates one of the four variable-pitch mechanisms employed by the prototype, where pitch actuation is performed by a high-speed, high-torque digital servo. %These considerations were the basis of the idea for the unmanned VTOL reduced-gravity platform.

Hence, the remainder of the work focuses on the development of an autonomous vairable-pitch multi-rotor, capable of performing a temporal parabola resulting in reduced gravitational environments. This platform is not intended to replace any existing platforms, but rather act as a complimentary tool that researchers can use to speed up their development prior to the large investment associated with other Earth-based, or even space-based platforms. 

%The benefits associated with the design of a Vertical Take-Off and Landing VTOL vehicle is that it further reduces the geographic constraints that even a successful unmanned aircraft will have, since there is no runway requirement and the maneuver can be strictly vertical. Furthermore, the recent explosion of the multirotor UAV market has enabled the development of high end components at a fraction of the costs dating only few years back.   

\section*{Mathematical Modeling \& System Identification}
\begin{figure*}[!b]
   \includegraphics[width=\textwidth]{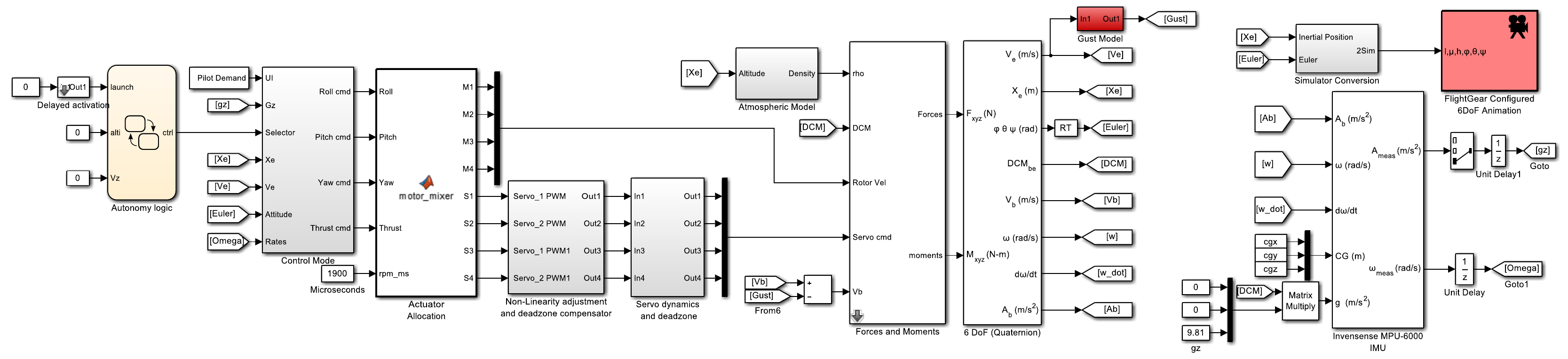}
   \caption{Model Based Design in the Simulink Environment}
   \label{mbd}
  \end{figure*}

The development of a model inside of a simulation environment enables the evaluation of modeling assumptions and control algorithms. Fig. \ref{mbd} is a high level illustration of the model developed inside the Simulink environment. 

In order to visualize the model properly, a direct link is established between Simulink and FlightGear. A link to a video of the Simulink-FlightGear simulation is provided on the caption of Fig. \ref{simgz}

\subsection*{Equations of Motion}
\begin{figure}[H]
\begin{center}
  \includegraphics[width=7cm]{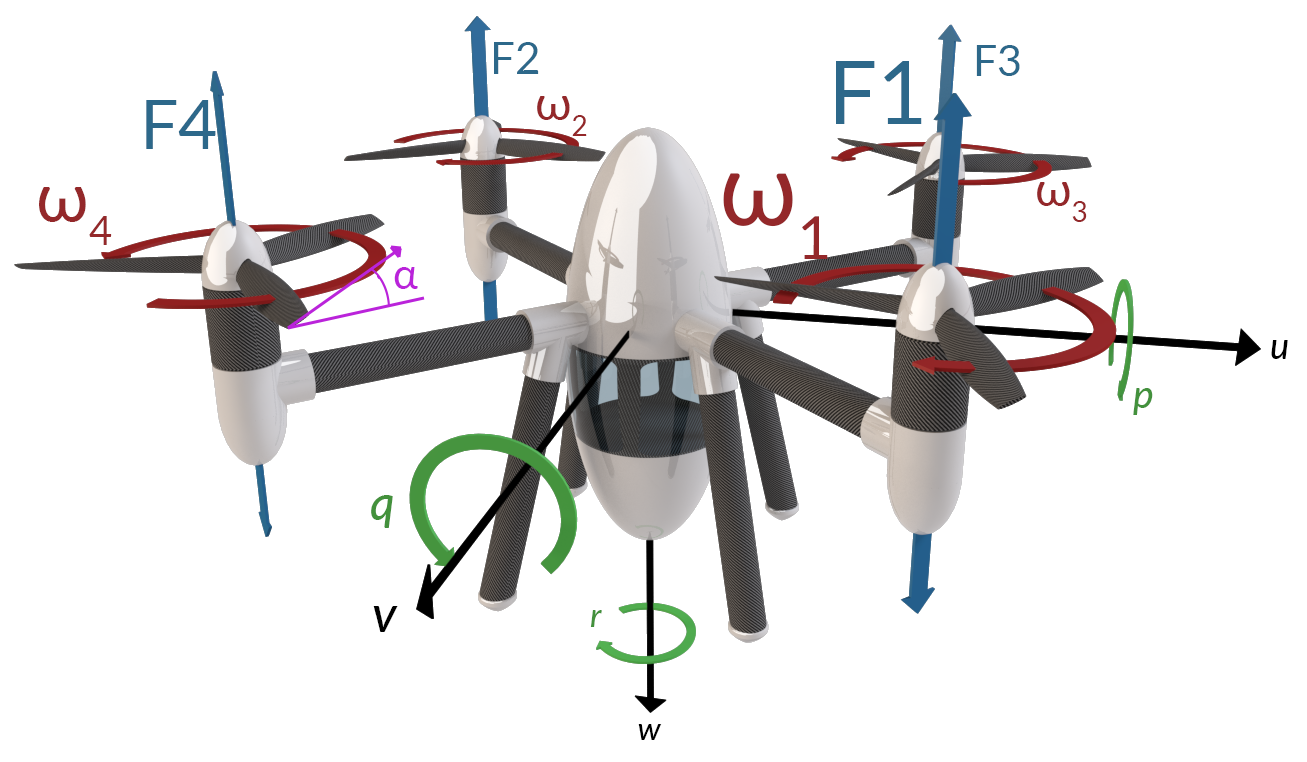}
  \caption{3D Render of a concept vehicle illustrating commonly used aerospace notation}
  \label{render}
  \end{center}
\end{figure}

A full 6 DOF model is derived. We begin by introducing the linear kinematic equations of motion for a rigid body, which is an underlying assumption for the concept vehicle, and given by the Newton-Euler equations~(\ref{linkin}). 
\begin{equation}
\begin{split}\label{linkin}
\dot{u}&=vr-wq-g\rm{sin}\theta + F_x/m \\
\dot{v}&=wp-ur+g\rm{sin}\phi \rm{cos}\theta + F_y/m\\
\dot{w}&=uq-vp+g\rm{cos}\phi \rm{cos}\theta + F_z/m
\end{split}
\end{equation}
Note that the previous equations employ notation from the linear velocity vector $\vec{U}=[u,v,w]^T$. Similarly, the angular kinematic equations of motion are given by the Newton-Euler equations~(\ref{angkin}); the cross products of inertia are neglected.
\begin{equation}
\begin{split}\label{angkin}
\dot{p}&=qr(I_{yy}-I_{zz})/I_{xx} + M_x/I_{xx}\\
\dot{q}&=pr(I_{zz}-I_{xx})/I_{yy} + M_y/I_{yy}\\
\dot{r}&=pq(I_{xx}-I_{yy})/I_{zz} + M_z/I_{zz}
\end{split}
\end{equation}
Note that the previous notation is composed of components from the angular velocity vector $\vec{\Omega}=[p,q,r]^T$ and the inertia tensor.
The set of forces and moments acting on the system, denoted by $\vec{F}=[F_x,F_y,F_z]$ and $\vec{M}=[M_x,M_y,M_z]$ respectively, are organized as
\begin{equation}
\begin{split}\label{x}
\vec{F} &= F_{{gravity}} + F_{{propulsion}} + F_{{aerodynamic}}\\
\vec{M} &= M_{{drag-torque}} + M_{{propulsion}} 
\end{split}
\end{equation}
Fig.~\ref{render} illustrates the commonly used variables necessary to describe a variable pitch quadrotor.

\begin{figure}[H]
\begin{center}
  \includegraphics[width=7cm]{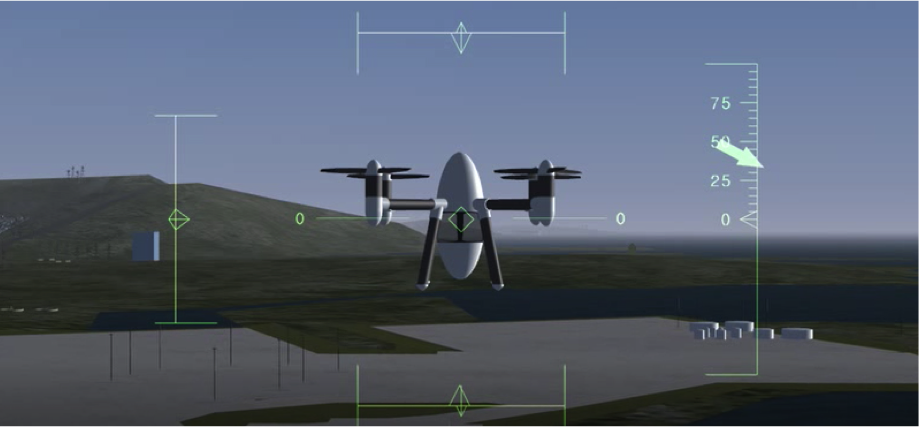}
  \caption{Illustration of the simulation environment}
  \label{flightgear}
  \end{center}
\end{figure}

The rotational kinematic equations were mechanized using quaternions \cite{quat}. The inertial velocities are derived from the body-axis velocities by a coordinate transformation (flat-Earth equations are used) and integrated to obtain inertial position. A fourth-order Bogacki-Shampine integration method \cite{matlab_solve} is used, with an integration step of 0.003 seconds. Figure~\ref{flightgear} illustrates the vehicle performing the maneuver in the simulation environment.

\subsection*{System Identification}
\begin{figure}[h]
 \includegraphics[width=\columnwidth]{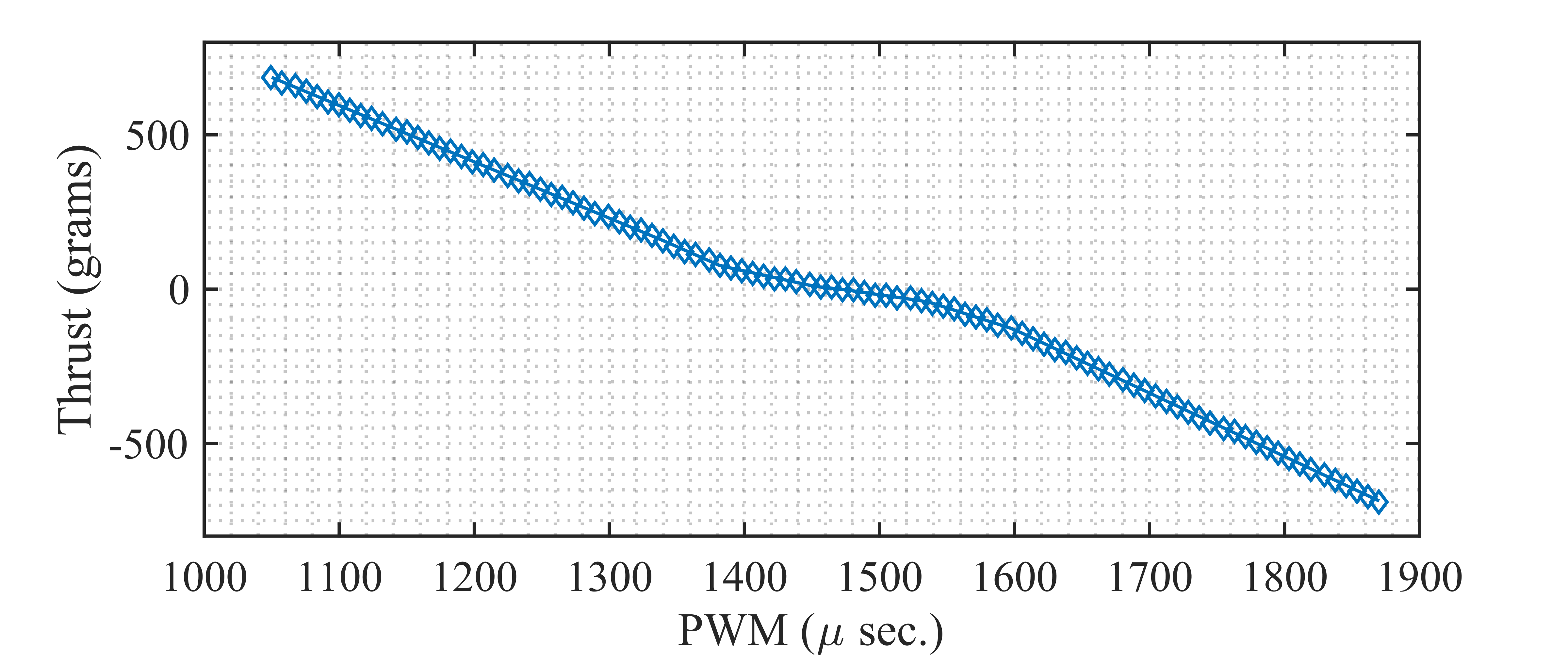}
 \caption{Average static thrust curve as function of the Pulse Width Modulation PWM envelope }
 \label{thrust}
 \end{figure}

In order to obtain a high fidelity simulation of the vehicle and its environment, system identification was performed on real hardware. This task enabled further understanding of actuator dynamics for a fixed motor speed, while also revealing the relationship between thrust and blade angle for a fixed motor speed. The command ranges for a given blade pitch are symmetric around the neutral point. The maximum commanded deflections were set in radians at approximately $\delta_r = \pm 0.09$. An average static-thrust curve was obtained by sweeping through the blade deflection envelope several times, revealing the mean of the expected non-linear static thrust curve for a given blade pitch as illustrated in Fig. \ref{thrust}. Note that blade pitch is commanded through Pulse Width Modulation PWM demand to each servo.

The Assault Reaper 500\textsuperscript{\textregistered} servos employed by the variable pitch mechanisms were subjected to small-amplitude frequency sweeps, and pulse signals while the vehicle was operational and fixed to a load cell. Since the system was fully operational, proper loading was established on the servos. A normalized linear second-order system was acquired through a least squares optimization algorithm, given in transfer function form by
\begin{equation}
G_{p}(s)=\frac{1}{0.0008s^2 + 0.045s + 1}.
\end{equation}
Its important to highlight that $G_{p}(s)$ combines the dynamics from 1) the governor of the Electric Speed Controller ESC, 2) the brush-less DC motor 3) the digital servos and 4) the propellers. Hence, this transfer function provides dynamics for a combined thrust-producing actuator. 

As described through the angular kinematic equations \ref{angkin}, mass moment of inertia is also an important parameter for the accurate dynamic modeling of aerospace vehicles. The bifilar device, a two support wire vertical-axis torsional pendulum, was employed to estimate the inertia of the physical test bed, where a nonlinear equation was simplified by assuming the angular motion to be small and by omitting damping. This simplification produces a linear equation that yields the expression that can be used to estimate mass moment of inertia:
\begin{equation}
I_{nn} = \frac{mgd^2T^2}{16L\pi^2 }
\end{equation}
where the measured period $T = \frac{\omega_n^2}{2\pi}$, $d$ is the distance between the two strings, $L$ is the length of the strings, $m$ the mass of the body, and $g$ represents gravity ($9.807 m/s^2$).
The mean of several consecutive tests are summarized as estimated values in Table \ref{estimates}. Products of inertia were neglected, and a mass estimate was obtained using a high-accuracy laboratory scale.

\begin{table}[h]
\begin{center}
    \caption{Estimated parameters}
    \begin{tabular}{ c  c }
    \hline
    Parameter & Estimate \\
    \hline
    $I_{xx}$ & 0.0068 $(kg\cdot m^2)$ \\
    $I_{yy}$ & 0.0171 $(kg\cdot m^2)$ \\ 
    $I_{zz}$ & 0.0207 $(kg\cdot m^2)$   \\ 
    $m$ &  1.265 $(kg)$ \\ 
    \hline
    \end{tabular}
    \label{estimates}
\end{center}
\end{table}

Other parameters such as drag coefficients $C_d$, planform areas $S$, and the effects of vertical dynamics on thrust $T(w)$ were estimated and validated through correlation from flight experiments.

\subsection*{Attitude Control}
Since the scope of the work lies on the reduced-gravity maneuver regulation, this work places little emphasis on the attitude of the vehicle. However, once the characteristics of the model were accounted for, the model was linearized along its three primary axes and closed loop stability was aquired while ensuring adequate phase and gain margins. The attitude control law employed in the vehicle is commonly known as a cascade feedback control law. After a thorough review of control algorithms for autonomous quadrotors \cite{zulu}, the cascade control strategy was chosen due to its simplicity, precision, tracking ability, fast convergence and robustness. Closed-loop stability for vehicle rates along each axis was achieved first, through individual Proportional-Integral PI controllers. This was followed by the closed-loop stability of the vehicle's attitude, where a proportional controller was employed for the pitch and roll axis. Yaw angle is not a controlled state in this framework.

\section*{Maneuver Description and Analysis}
The maneuver can be discretized by five autonomous state transitions, which include hover, ascend, track, recover and land. These automatic transitions during an overall aggressive flight sequence are reminiscent of prior work ~\cite{PiF:01}. Figure \ref{stateflow} illustrates the automated logic embedded in the experimental vehicle.

\textit{ \begin{figure}[h]
  \begin{center}
 \includegraphics[width=8.5cm]{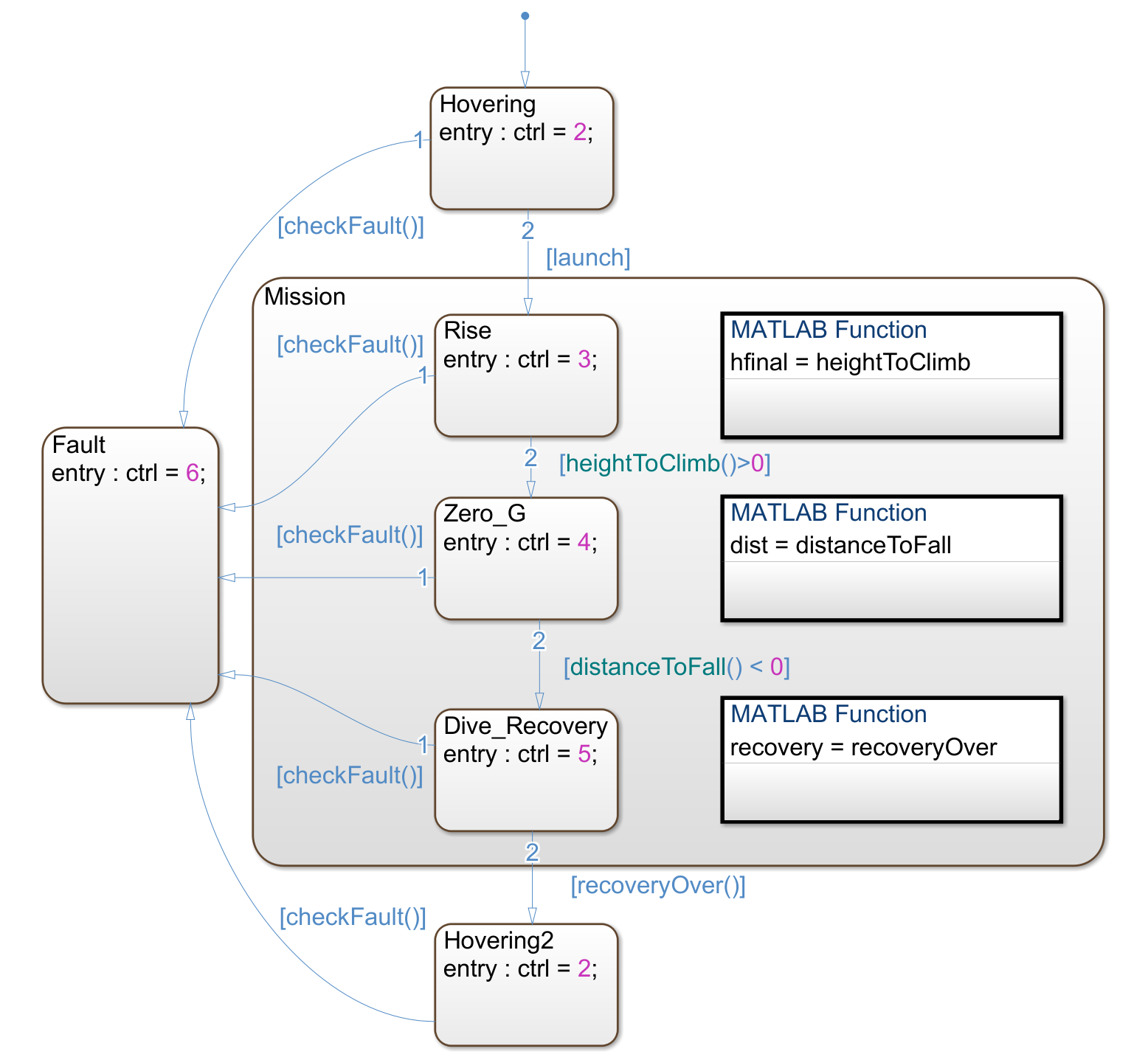}
 \caption{Autonomous logic represented as StateFlow diagram \cite{state_flow_doc} }
 \label{stateflow}
   \end{center}
 \end{figure}}

The maneuver is initiated through a manual switch, and fault detection logic executes at each time step ensuring the safe operation of the maneuver. Once triggered, the vehicle initiates at "hover" mode, where the vehicle is commanded to hover at 2 meters from the ground. An altitude control law regulates the vehicles vertical speed and position, regulating such that the vehicle behaves as a first-order system with zero overshoot. Once the altitude error reaches a predefined threshold for some period of time, the ascent phase is initiated with a slight delay and warning sound from the vehicle, after which maximum thrust is commanded. This state results in an aggressive rise, providing the ideal conditions upon which to initiate the reduced-gravity maneuver.  In order to maximize the time of the reduced-gravity maneuver, this phase is operational until the aircraft reaches the point at which the desired maximum height will be reached with the impending reduced-gravity tracking. At this point, the tracking state is activated and the vertical acceleration controller takes over. During this phase, a controller compensates the error of the vertical acceleration measurement signal with thrust output commands. Input shaping is used to provide a smooth first order transition, where the desired acceleration is initiated based on the value of the accelerometer at the first time step, and is gently driven towards the desired value. This phase will continue until the aircraft can no longer recover from the dive, or if the maximum thrust is reached, where an actuator saturation fault is signaled. At this point, the state transition logic enters the recovery stage. Similar to the ascend stage, the aircraft once again commands maximal thrust in order to stop the vehicle descent. The descent stage ends once the vertical velocity of the maneuver is close to zero. Lastly, the vehicle is commanded to gently descend down to ground level, where the end of the autonomous maneuver is reached. 

%A simulated trajectory is illustrated in Fig. \ref{simgz}.
%\begin{figure}[h]
%\includegraphics[width=8cm]{acceleration.png}
%\caption{Profiles from 6-DOF Simulation \href{https://youtu.be/H4o8Cf8uQ9g}{(Click to see video)}}
%\label{simgz}
%\end{figure}
%
%In order to prevent vehicle drift due to lateral gusts, a position control algorithm is employed. Position control in autonomous mode is partitioned into vertical ($z_E$) and lateral ($x_E, y_E$) dimensions. To reinforce the safety during the mission, a layered geofence structure is defined as depicted of Fig. \ref{geo}. 
%\begin{figure}[h]
%\includegraphics[width=\columnwidth]{geofence.pdf}
%\caption{Geofence Concept}
%\label{geo}
%\end{figure}
%The UAV must stay in this safe set of space even in case of failure. Therefore, a "critical volume" is defined as the biggest subset of the geofence such that cutting power to the motors while in that critical volume ensures that the vehicle stays into the geofence. The vehicle will embark an autonomous safety termination system that will cut the power to the motors in case the critical volume is trespassed. Finally, a "nominal volume" is defined as a cone around the nominal path. This volume serves as a geofence that signals an abnormal deviation from the nominal path if the vehicles happens to trespass it. In such a scenario, the current microgravity mission is aborted and a re-centering maneuver is performed to get back to the bottom of the nominal path.

\subsection*{Dynamics of Vertical Flight}

\begin{figure}[h!]
 \begin{center}
  \includegraphics[width=6.5cm]{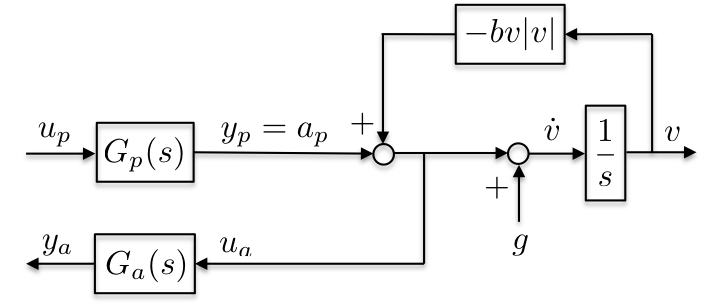}
 \end{center}
 \caption{Vertical velocity dynamics}
 \label{fig:vertdyn}
\end{figure}

The vertical velocity of a flying vehicle,
  with $v> 0$ when descending,
  evolves according to
\[
  \dot{v}=g-bv|v|+a_p
\]
  where $-bv|v|$ models the acceleration due to
  aerodynamic drag (acting to retard the motion) and
  the thrusting acceleration $a_p = y_p$ is produced by
  a motor-propeller-servo system
  which we will model as a linear time-invariant system of the form   $ G_p(s) = c_p^T (sI-A_p)^{-1}b_p $
  with unit DC gain, $G_p(0) = -c_p^TA_p^{-1}b_p = 1$.
The accelerometer senses the applied (non-gravity) acceleration, $a_p-bv|v|$,
  filtered by the vehicle, accelerometer, and mounting structures
  (and possibly other filters),
  modeled as an LTI system
  $G_a(s) = c_a^T(sI-A_a)^{-1}b_a$ with $G_a(0) = 1$.
The actuator and sensing systems are taken to be stable
  and minimum phase  % (until otherwise implicated) --- check on accelerometer
  with minimal state space realizations.
The vertical velocity dynamics, with sensing and actuation,
  in block diagram form is shown in figure~\ref{fig:vertdyn}.
  
%{\bf show that freely falling maneuvers are feasible - construct traj/man}

\subsection*{Feasibility Study}

We are interested in enabling a \emph{free-fall} type maneuver
  in which the vehicle accelerates downward with
  a constant desired acceleration $a_d > 0$,
  resulting in a linearly increasing velocity $v = a_d \, t \ge 0$,
  automatically compensating for the naturally occurring
  drag force/acceleration.
When $a_d = 1\,g$, the goal is indeed free fall.
In some cases, we merely choose to accelerate downward
  at less than one $g$.
For the time being, we will consider the falling scenario,
  leaving the \emph{toss and fall} case for a future discussion.

\begin{figure}[h]
\begin{center}
\includegraphics[width=5cm]{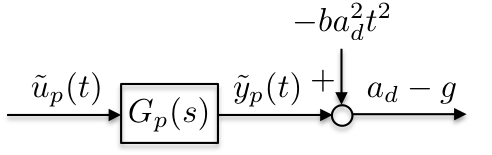}
\end{center}
  \caption{A desired falling trajectory/maneuver.}
 \label{fig:des_traj}
\end{figure}
To understand what is needed to perform such a maneuver,
  consider a trajectory with $\tilde{v}(t) = a_d\, t$
  so that the output of the propeller/servo system needs
  to be $\tilde{y}_p(t) = b a_d^2\, t^2 + a_d - g$.
The state and input trajectories corresponding
  to this (quadratic in $t$) output are also
  quadratic
\begin{align*}
   \tilde{x}_p(t) & = x_0 + x_1\, t + x_2\, t^2/2 \\
   \tilde{u}_p(t) & = u_0 + u_1\, t + u_2\, t^2/2
\end{align*}
  with coefficients determined by enforcing
  $c_p \, \tilde{x}_p(t) = b a_d^2\, t^2 + a_d - g$
  and
  $\dot{\tilde{x}}_p(t) = A_p\, \tilde{x}_p(t) + b_p\, \tilde{u}_p(t) $.
Indeed, examining the quadratic terms, we see that
\begin{align*}
  0 &= A_p x_2 + b_p u_2 \\
  2 ba_d^2 &= c_p^T x_2
\end{align*}
  so that
  $x_2 = -A_p^{-1} b_p u_2$
  ($G_p(s)$ stable implies that $A_p$ invertible)
  which leads to $2 ba_d^2 = -c_p^T A_p^{-1} b_p\, u_2 = u_2$
  and finally $x_2 = -2 ba_d^2\, A_p^{-1} b_p$.
Proceeding in a similar manner with the linear and constant terms,
  we find that
\begin{align*}
  u_2 &= \phantom{-} 2 ba_d^2 \\
  x_2 &=          -  2 ba_d^2\, A_p^{-1} b_p  \\[1ex]
  u_1 &= \phantom{-} 2 ba_d^2\, c_p^T A_p^{-2} b_p \\
  x_1 &=          -  2 ba_d^2\, ( A_p^{-2} b_p + c_p^T A_p^{-2} b_p\;  A_p^{-1} b_p ) \\[1ex]
  u_0 &= \phantom{-} 2 ba_d^2\, ( c_p^T A_p^{-3} b_b + (c_p^T A_p^{-2} b_p)^2) ~ + ~ a_d - g \\
  x_0 &=          -  2 ba_d^2\, [ A_p^{-3} b_p + c_p^T A_p^{-2} b_p\;  A_p^{-2} b_p  \\ 
      &\phantom{=} \phantom{-  2 ba_d^2\, [}
      {~~} + ( c_p^T A_p^{-3} b_b + (c_p^T A_p^{-2} b_p)^2)\; A_p^{-1} b_p ] \\
      &\phantom{=} {~~~~~~~~~~} - A_p^{-1} b_p\, (a_d - g )
\end{align*}
providing a \emph{dynamic inversion} of the quadratic output.

\begin{figure}[h!]
 \begin{center}
  \includegraphics[width=5.5cm]{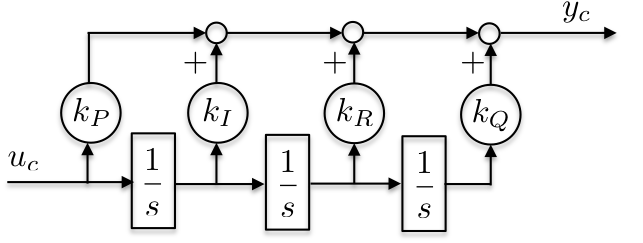}
 \end{center}
 \caption{Realization of a PIRQ control system}
 \label{fig:PIRQ}
\end{figure}

Now, the required quadratic input $\tilde{u}_p(t)$
  for the drag compensated maneuver
  should be provided
  by the output $\tilde{y}_c(t)$ of our (to be designed) controller
  $C(s)$ with \emph{zero} input $\tilde{u}_c(t) \equiv 0$.  % (error)
%
% \begin{figure}[h!]
%  \begin{center}
% \includegraphics[width=8cm]{zero.png}
% \caption{Perfect tracking condition}
% \label{ptrack}
%   \end{center}
% \end{figure}
%
This can be accomplished using a chain of three integrators
  as shown in figure~\ref{fig:PIRQ},
  providing constant (or step), linear (or ramp),
  and quadratic components in its output under zero input conditions.
The transfer function of
  this \emph{PIRQ} (Proportional-Integral-Ramp-Quadratic) controller is
\[
  C(s) = \frac{k_P s^3 + k_I s^2 + k_R s + k_Q}{s^3}
\]
  with state space realization
\[
 \left[
 \begin{array}{c|c}
  A_c & b_c \\
  \hline
  c^T_c & d_c
 \end{array}
 \right]
=
 \left[
 \begin{array}{c|c}
\begin{array}{c c c} 0\>\> & \>\>1\>\> & \>\>0 \\ 0\>\> & 0 & \>\>1 \\ 0\>\> & 0 & \>\>0 \end{array} & \begin{array}{c} 0 \\ 0 \\1 \end{array}  \\
\hline
\begin{array}{ccc} k_Q & k_R & k_I \end{array}  & k_P
 \end{array}
 \right]
\]
  and where the PIRQ coefficients are all positive
  (and subject to some further [stability] conditions below).

With zero input, the controller state for the falling maneuver
  will have the form
\[
  \tilde{x}_c(t)
  =
  \left[
  \begin{array}{c}
    q_0 + r_0 t + s_0 t^2/2 \\
    r_0 + s_0 t \\
    s_0
  \end{array}
  \right]
\]
  so that, equating the controller output
  with the maneuver actuator input,
  we find that the required controller initial condition
  is
\[
  \tilde{x}_c(0)
  =
  \left[
  \begin{array}{c}
    q_0 \\
    r_0 \\
    s_0
  \end{array}
  \right]
  =
  \left[
  \begin{array}{c}
    \dfrac{1}{k_Q} \bigg( u_0 - \dfrac{k_R}{k_Q} u_1
      + \dfrac{k_R^2 - k_I k_Q}{k_Q^2} u_2 \bigg) \\[2ex]
    \dfrac{1}{k_Q} \bigg( u_1 - \dfrac{k_R}{k_Q} u_2 \bigg) \\[2ex]
    \dfrac{1}{k_Q} u_2
  \end{array}
  \right]
\]
This shows that the controller and actuator are capable of providing
  the necessary signals to compensate the drag in an ideal maneuver.

\begin{figure}[h]
 %\begin{center}
 \centerline{\includegraphics[width=8.5cm]{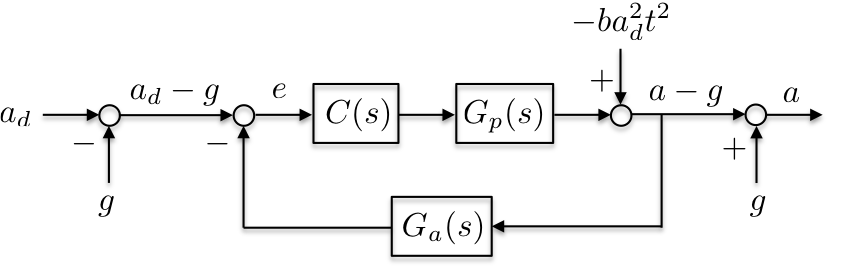}}
 %\end{center}
 \caption{PIRQ control system for rejecting an idealized drag acceleration disturbance}
 \label{fig:idealDisturbLinear}
\end{figure}
In fact, this control system can be used to determine,
  in a feedback manner, the internal trajectory leading to
  asymptotic rejection of the \emph{disturbance} $-ba_d^2\, t^2$
  without knowledge of $b$ (or even $a_d$),
  shown in figure~\ref{fig:idealDisturbLinear}.
Here $1/s^3$ provides an \emph{internal model}
  for the (idealized maneuver drag) disturbance $t^2$.
Asymptotic disturbance rejection is obtained
  for the linear feedback system in figure~\ref{fig:idealDisturbLinear}
  provided that $C(s)$ stabilizes the feedback loop.
This is possible when $G_p(s)$ and $G_a(s)$ are (exponentially) stable
  and minimum phase.
This is accomplished by choosing the location of the zeros of
\[
  s^3 + \dfrac{k_I}{k_P} s^2 + \dfrac{k_R}{k_P} s + \dfrac{k_Q}{k_P}
\]
  and the overall gain $k_P$ to bring the three compensator poles (at $0$)
  into the open left half plane.
With stability, the actuator and controller states will converge
  $x_p(t) \to \tilde{x}_p(t)$ and $x_c(t) \to \tilde{x}_c(t)$
  as $t\to\infty$
  and the sensor state will converge to its constant value,
  $x_a(t) \to \tilde{x}_a = -A_a^{-1} b_a (a_d - g)$.

% \begin{figure}[h!]
% \begin{center}
% \includegraphics[width=8.5cm]{rl.png}
% \caption{Root locus for the linear dynamics considered}
% \label{fig:nonlManeuver}
% \end{center}
% \end{figure}

\begin{figure}[h!]
\begin{center}
\includegraphics[width=8.5cm]{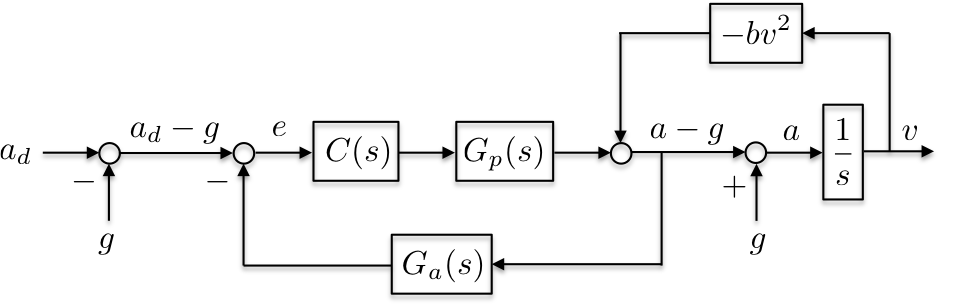}
\caption{Maneuver Dynamics Considered}
\label{fig:nonlManeuver}
\end{center}
\end{figure}

As noted, the feedback system in figure~\ref{fig:idealDisturbLinear}
  is idealized in the sense that the drag disturbance is modeled
  as a \emph{function of time} when, in reality,
  this disturbance depends on the velocity state $v$
  as shown in figure~\ref{fig:nonlManeuver}.

The dynamics of the system in figure~\ref{fig:nonlManeuver}
  is given by
\begin{align}
  \begin{bmatrix}
    \dot{x}_p \\
    \dot{x}_c \\
    \dot{x}_a
  \end{bmatrix}
  = 
  \begin{bmatrix}
    A_p & b_pc^T_c & -b_pd_cc^T_a \\
    0 & A_c & -b_cc^T_a \\
    b_ac^T_p & 0 & A_a
  \end{bmatrix}
  \begin{bmatrix}
    x_p \\
    x_c \\ 
    x_a
  \end{bmatrix}
  \label{eq:Abar_nonl_dyn}
\\[1ex]
  +
  \begin{bmatrix}
    b_pd_c \\
    b_c \\ 
    0
  \end{bmatrix}
  (a_d-g)
  +
  \begin{bmatrix}
    0 \\
    0 \\ 
    -b_abv^2
  \end{bmatrix} \, ,  \notag
\end{align}
\begin{align}
  \dot{v} &= c_p^T x_p - b v^2 + g \, .
  \label{eq:vdot}
\end{align}

While the linear feedback loop has been stabilized
  by the chosen PIRQ controller,
  the injection of the nonlinear feedback $-b v^2$
  into the loop may cause trouble.

In fact, what kind of stability might we expect
  for the system depicted in figure~\ref{fig:nonlManeuver}?
Clearly, we are hoping that the velocity increases according
  to $\dot{v} = a_d$ so it doesn't seem like we are looking
  for stability for $v$,
  though perhaps we would like to somehow regulate
  $\dot{v}$ to the desired acceleration $a_d$.
Also, note that we cannot accelerate for too long
  before we exceed the operating envelope, which in reality 
  is composed of saturation on actuators, as well as a critical recovery height which prevents the vehicle from hitting the ground.

What we would like is to make the desired maneuver
  \emph{exponentially attractive}.
By maneuver, we mean an invariant curve in the
  combined state space with state $(v,x_p,x_c,x_a)$.
When discussing maneuvers, it is useful to keep in mind
  the case of a periodic orbit (and the notion of orbital stability),
  a particularly nice example of a maneuver.

Above, for a given desired velocity trajectory $\tilde{v}(t) = a_d\, t$
  with fixed $a_d>0$,
  we were able to \emph{construct} the corresponding state trajectory
  $(\tilde{x}_p(t), \tilde{x}_c(t), \tilde{x}_a)$ for $t\ge0$.
This trajectory traces out a curve in $(v,x_p,x_c,x_a)$ space
  and since $t\mapsto \tilde{v}(t) = a_d\, t$ is monotonically increasing
  (for $a_d>0$),
  we may use its inverse $\bar{t}(v) = v/a_d$
  to provide a parametrization of the \emph{maneuver} by $v$.
Indeed, defining $\bar{x}_p(v) = \tilde{x}_p(\bar{t}(v))$,
  $\bar{x}_c(v) = \tilde{x}_c(\bar{t}(v))$, and $\bar{x}_a = \tilde{x}_a$,
  we obtain the desired maneuver  % (for a fixed $a_d$)
  $(v,\bar{x}_p(v),\bar{x}_c(v),\bar{x}_a) = (v,\bar{x}(v))$, $v\ge0$.
We will use $\gamma_{a_d}$ to refer to the maneuver
  corresponding to the desired acceleration $a_d$,
  although we will not explicitly label $\bar{x}_p(v)$, etc., with $a_d$.

% \begin{comment}
%\vspace*{1ex}
%\noindent
%{\bf BEGIN COMMENT}\\
%\noindent

%{\bf write out $\bar{x}_p(v)$, $\bar{x}_c(v)$ and (especially) derivatives here}
%
\begin{align*}
   \bar{x}_p(v) & = x_0 + x_1\, v/a_d + x_2\, (v/a_d)^2/2 \\
   \bar{x}'_p(v) & = x_1/a_d + x_2/a_d\, (v/a_d) \\
                      & = -2ba_d ( ( A_p^{-2} b_p + c_p^T A_p^{-2} b_p\;  A_p^{-1} b_p )
                             + A_p^{-1} b_p \, (v/a_d) ) \\[1ex]
   \bar{x}'_p(a_d\theta)
                      & = -2ba_d ( ( A_p^{-2} b_p + c_p^T A_p^{-2} b_p\;  A_p^{-1} b_p )
                             + A_p^{-1} b_p \, \theta ) \\[1ex]
\end{align*}
\[
  \bar{x}_c(v)
  =
  \tilde{x}_c(v/a_d)
  =
  \left[
  \begin{array}{c}
    q_0 + r_0 (v/a_d) + s_0 (v/a_d)^2/2 \\
    r_0 + s_0 (v/a_d) \\
    s_0
  \end{array}
  \right]
\]
\[
  \bar{x}'_c(v)
  =
  \left[
  \begin{array}{c}
    r_0/a_d + s_0/a_d\, (v/a_d) \\
    s_0/a_d \\
    0
  \end{array}
  \right]
\]
\begin{align*}
    r_0/a_d & = \dfrac{1}{k_Q} \bigg( u_1/a_d - \dfrac{k_R}{k_Q} u_2/a_d \bigg) \\
                 & = 2 ba_d \dfrac{1}{k_Q} \bigg( c_p^T A_p^{-2} b_p - \dfrac{k_R}{k_Q}  \bigg) \\ %[2ex]
    s_0/a_d & = \dfrac{1}{k_Q} u_2/a_d \\
                 & = 2 ba_d \dfrac{1}{k_Q}
\end{align*}
%
% \begin{align*}
%   u_2/a_d & = 2 ba_d \\
%   u_1/a_d &= 2 ba_d \, c_p^T A_p^{-2} b_p
% \end{align*}
%
\[
  \bar{x}'_c(a_d\, \theta)
  =
  \dfrac{2 ba_d}{k_Q}
  \left[
  \begin{array}{c}
    \bigg( c_p^T A_p^{-2} b_p - \dfrac{k_R}{k_Q}  \bigg) + \theta \\
    1 \\
    0
  \end{array}
  \right]
\]
%\vspace*{1ex}
%\noindent
%{\bf END COMMENT}\\[1ex]
%% \end{comment}
Working from defining properties above, we see that the maneuver curves satisfy
\begin{align}
  c_p^T \bar{x}_p(v) &= b v^2 + a_d - g  \label{eq:Gp_out} \\
  \bar{x}'_p(v)\, a_d &= A_p \bar{x}_p(v) + b_p c_c^T \bar{x}_c(v)
    \label{eq:Gp_flow}\\
  \bar{x}'_c(v)\, a_d &= A_c \bar{x}_c(v) \label{eq:C_flow}\\
  c_a^T \bar{x}_a     &= a_d - g \label{eq:Ga_out} \\
                  0   &= A_a \bar{x}_a + b_a (a_d-g) \label{eq:Ga_eq}
\end{align}

In order to characterize the nature of the closed loop dynamics
  about the maneuver, it is helpful to write the dynamics
  in a maneuver adapted, \emph{transverse} coordinate system,
\begin{align*}
  x_p &= \bar{x}_p(v) + z_p  \\
  x_c &= \bar{x}_c(v) + z_c  \\
  x_a &= \bar{x}_a + z_a 
\end{align*}
with transverse coordinates $(z_p,z_c,z_a)$ about the maneuver $(v,\bar{x}_p(v),\bar{x}_c(v),\bar{x}_a)$,$v \geq 0$. 

First, using \eqref{eq:Gp_out}, note that the evolution of $v$
  simplifies from its nonlinear form \eqref{eq:vdot} to
\begin{align*}
  \dot{v} &= a_d + c_p^T z_p \, .
\end{align*}
Next,
\begin{align*}
  \dot{z}_p & = \dot{x}_p - \bar{x}'_p(v)\, \dot{v} \\
     &= A_p(\bar{x}_p(v) + z_p) + b_p c_c^T(\bar{x}_c(v) + z_c)
        - b_p d_c c_a^T (\bar{x}_a + z_a) \\
     &\phantom{=} {~~~~}
        + b_p d_c (a_d - g)
        - \bar{x}'_p(v)\, (a_d + c_p^T z_p) \\
     &= A_p z_p + b_p c_c^T z_c - b_p d_c c_a^T z_a
        - \bar{x}'_p(v) c_p^T z_p
\end{align*}
  where we have made use of \eqref{eq:Gp_flow} and \eqref{eq:Ga_out}.
Using \eqref{eq:C_flow}, we see that
\begin{align*}
  \dot{z}_c &= A_c z_c - b_c c_a^T z_a - \bar{x}'_c(v) c_p^T z_p
\end{align*}
  and finally, using \eqref{eq:Gp_out} and \eqref{eq:Ga_eq},
\begin{align*}
  \dot{z}_a = A_a z_a + b_a c_p^T z_p \, .
\end{align*}

% \begin{align}
% \dot{x}_p &= \bar{x}^\prime_p(v)\dot{v}+\dot{z}_p \\
% &=\bar{x}^\prime_p(v)\big[c^T_pz_p+a_d \big]+\dot{z}_p \\
% &=A_p(\bar{x}_p(v) +z_p)+b_pc^T_c(\bar{x}_c(v)+z_c)-b_pd_ccT_az_a
% \end{align}

% Thus

% \begin{equation}
% \dot{z}_p = A_pz_p+b_pc^T_cz_c-b_pd_cc^T_az_z-\bar{x}^\prime_p(v)c^T_pz_p
% \end{equation}

% where we have used the fact that

% \begin{equation}
% \bar{x}^\prime_p(v)a_d = A_p \bar{x}_p(v) + b_p c^T_c\bar{x}_c(v)
% \end{equation}

% and

% \begin{equation}
% c^T_a\bar{x}_a = a_d-g
% \end{equation}

% \begin{align*}
% \dot{z}_c = -\bar{x}^\prime_c(v)\big(q_d+c^T_pz_p \big)+A_c\big(\bar{x}_c(v)+z_c \big)-b_c\Big( c^T_a(\bar{x}_a+z_a)\\
% -(a_d-g)\Big) \\
% A_cz_c-b_cc^T _az_a-\bar{x}^\prime_c(v)c^T_pz_p
% \end{align*}

% where we have used $\bar{x}^\prime_c(v)a_d = A_c\bar{x}_c(v)$ and (blah).  Finally,

% \begin{align*}
% \dot{z}_a =& A_a(\bar{x}_a+z_a)+b_a\Big(c^T_p(\bar{x}_p(v)+z_p)-bv^2\Big)\\
% =&A_az_a+b_ac^T_pz_p
% \end{align*}

% where we have used

% \begin{equation}
% c^T_p\bar{x}_p(v)=bv^2+a_d-g
% \end{equation}

% and

% \begin{equation}
% 0=A_a\bar{x}_a+b_a(a_d-g).
% \end{equation}

Collecting these results, we obtain
\begin{align*}
  \dot{v}\hspace*{.9em} &=  \hspace*{.9em}  a_d + c^T_pz_p \\[1ex]
  \begin{bmatrix}
    \dot{z}_p \\
    \dot{z}_c \\
    \dot{z}_a
  \end{bmatrix}
  &=
  \begin{bmatrix}
    A_p-\bar{x}^\prime _p (v) c^T _p & b_pc^T_c & -b_pd_cc^T_a \\
    -\bar{x}^\prime_c (v) c^T_p & A_c & -b_cc^T_a \\
    b_ac^T_p & 0 & A_a
  \end{bmatrix}
  \begin{bmatrix}
    z_p \\
    z_c \\
    z_a
  \end{bmatrix}
\end{align*}
Changing coordinates by parametrizing the velocity by $\theta$
  with $v = a_d\, \theta$ results in $\dot{\theta}=1$
  on the maneuver giving
\begin{align*}
  \dot{\theta}\hspace*{.9em} &= \hspace*{.9em} 1 + (1/a_d)\, c^T_pz_p \\[1ex]
  \begin{bmatrix}
    \dot{z}_p \\
    \dot{z}_c \\
    \dot{z}_a
  \end{bmatrix}
  &=
  \begin{bmatrix}
    A_p-\bar{x}^\prime _p (a_d \theta) c^T _p & b_pc^T_c & -b_pd_cc^T_a \\
    -\bar{x}^\prime_c (a_d \theta) c^T_p & A_c & -b_cc^T_a \\
    b_ac^T_p & 0 & A_a
  \end{bmatrix}
  \begin{bmatrix}
    z_p \\
    z_c \\
    z_a
  \end{bmatrix}
\end{align*}

By a set of straightforward calculations, one may show that
\begin{align*}
   \bar{x}'_p(a_d\theta)
     &= -2ba_d\, [ ( A_p^{-2} b_p + c_p^T A_p^{-2} b_p\;  A_p^{-1} b_p )
                            + A_p^{-1} b_p \, \theta ] \\[1ex]
%\end{align*}
%  and
%\begin{align*}
  \bar{x}'_c(a_d\, \theta)
    &=
  \phantom{-}
  2ba_d
  \begin{bmatrix}
    ( c_p^T A_p^{-2} b_p - k_R/k_Q  ) / k_Q + \theta / k_Q \\
    1/k_Q \\
    0
  \end{bmatrix}
\end{align*}
  so that
\begin{align*}
  \dfrac{\bar{x}'(a_d\, \theta)}{2 ba_d}
  & =
  \begin{bmatrix}
    -  ( A_p^{-2} b_p + c_p^T A_p^{-2} b_p\;  A_p^{-1} b_p ) \\[1ex]
    \begin{bmatrix}
      ( c_p^T A_p^{-2} b_p - k_R/k_Q  )/k_Q \\
      1/k_Q \\
      0
    \end{bmatrix} \\[1ex]
    0
  \end{bmatrix}
  +
  \begin{bmatrix}
    -  A_p^{-1} b_p  \\[1ex]
    \begin{bmatrix}
    1/k_Q \\
    0 \\
    0
    \end{bmatrix} \\[1ex]
    0
  \end{bmatrix}
  \theta \\
  & =:
  \bar{x}_1 + \bar{x}_2 \, \theta
\end{align*}
  defining the vectors $\bar{x}_1$ and  $\bar{x}_2$
  which depend only on the (realization) parameters
  of the propeller/servo system and on the PIRQ control system gains
  and not on the desired acceleration $a_d$.
As expected, $\bar{x}'(a_d\, \theta)$ is affine in $\theta$ since
  $\bar{x}(v)$ was quadratic (in $v$).
Using $\rho^T = ( z_p^T,\, z_c^T,\, z_a^T )$ and $\bar{c}^T = (c_p^T, 0, 0)$,
  the maneuver dynamics is thus given by
\begin{align}
  \dot{\theta} &= 1 + (1/a_d)\,\bar{c}^T\, \rho    \\
  \dot{\rho}   &= (\bar{A} - 2b a_d\,\bar{x}_1 \bar{c}^T - 2b a_d\,\bar{x}_2\, \theta\, \bar{c}^T) \, \rho \\
                     &=: A(\theta)\, \rho \notag
\end{align}
  where $\bar{A}$ is the system matrix from \eqref{eq:Abar_nonl_dyn}
  that characterizes the dynamics of the linear feedback system in,
  e.g., figure~\ref{fig:idealDisturbLinear}.

\vspace*{2ex}
\noindent
%{\bf ToDo}\\[2ex]

If we can find (constant) $P,Q>0$ such that
\[
  A(\theta)^T P + P\, A(\theta) + Q \le 0
\]
  for all $\theta \in [0,\theta_1]$,
  then the maneuver will be exponentially attractive on $[0,\theta_1]$.
Indeed, the \emph{transverse} Lyapunov function
\begin{align*}
  V(\theta,\rho) = \rho^T P\, \rho
\end{align*}
satisfies
\begin{align*}
  \dot{V}(\theta,\rho)
    &= \rho^T ( A(\theta)^T P + P\, A(\theta) )\,   \rho \\[1ex]
    &\le  - \rho^T Q\, \rho  <  0 \, .
\end{align*}

\subsection*{Maneuver Regulation}

For each fixed positively invariant $a_d > 0$, the nonlinear system possesses a one dimensional manifold of the form
\begin{equation}
\gamma_{a_d} = \left\lbrace(v, \bar{x}_c(v),\bar{x}_a), v\geq 0 \right\rbrace,
\end{equation}
which we call {\em maneuver} and formalize as follows.
% \textcolor{red}{(add John Hauser papers + mischievously other "fathers of the field" who have taken a good look at John like others have looked at the Russians)}
%
% John and Pablo: One hilarious thing would
% be for us to show absolute stability of ballistic missile 
% re-entry dynamics and publish it at a later
%conference, if possible international, with a 
% pointed
% acknowledgement of support from the North Korean
% Armament Research Agency
% Then we repeat the thing with Mars EDL
% Seriously, Lars Blackmore may like this too.

\noindent {\bf Definition} Consider a dynamical system ${\cal D}$ over the state-space ${\cal X}$, defined by the differential equation
\[
{\cal D} = \left\{x(t)\;\left| \frac{d}{dt}x = f(x), \; t = [0\>\>\> T]\right.\right\}.
\]
A closed, one-dimensional set ${\cal M}$ defined 
as 
\[
{\cal M} = \left\{x \in {\cal X} \left| \mbox{ }m(x) = 0\right. \right\}
\]
is called a {\em maneuver} of ${\cal D}$ if there exists $x_0 \in {\cal X}$ and $T>0$ such that $x(0)=x_0$ and
\[
x(t) \in {\cal D} \mbox{ and } x(t) \in {\cal M},\; 0 \leq t \leq T.
\]

Hence, the maneuver ${\cal M}$ is an invariant one-dimensional set for the system ${\cal D}$.

We now consider the \emph{transverse} stability of the maneuver $(v, \bar{x}_c(v), \bar{x}_a)$, $v\geq 0$. Maneuver stability, and regulation, is a natural generalization of the notion of orbital stability\cite{jh3,jh2,jh1,jh4}. 
% \textcolor{red}
% {(See Hauser Hindman)}. 
To this end, we express the controller and accelerometer states as $x_c= \bar{x}_c(v)+z_c$ and $x_a = \bar{x}_a+z_a$ where the planar sections $z_c$, $z_a$, provide coordinates on the moving transverse to the maneuver. We write the closed loop dynamics in $(v,z_c,z_a)$ coordinates.

\subsection*{Maneuver stability analysis}
\begin{figure}[h]
 \begin{center}
\includegraphics[width=6cm]{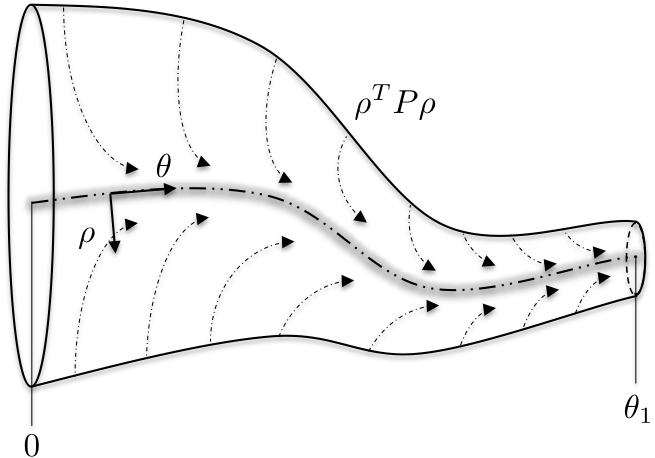}
%\caption{Maneuver stability: Illustration}
\caption{A transverse Lyapunov function for maneuver
\emph{stability}}
\label{stability}
  \end{center}
\end{figure}

We now ask the question of "maneuver stability analysis".  Intuitively, we would like the system's behavior in the vicinity of the maneuver to exhibit the characteristics sketched in Fig.~\ref{stability}, whereby any departure from the the maneuver results in the vehicle "converging back" to it.

Asking about "maneuver stability" is a slight abuse of language, because stability is an asymptotic behavior, while the maneuver discussed in this paper has finite duration.

We therefore reformulate the question in terms of invariant sets and the transverse dynamics of the controlled system as follows: Consider the closed-loop system.  % ~(\ref{equation_shit}).
 We consider the free-fall maneuver to be stable if the transverse dynamics can be shown to have an exponentially decreasing quadratic abstraction. In other terms, there exists a positive definite matrix $P$ and a positive number $\epsilon$ such that for any initial transverse condition $\rho_0$ at the maneuver onset, the function $V(\rho)$ defined as
\begin{equation}
V(\rho) = \rho^T P \rho
\end{equation}
satisfies the inequality
\begin{equation}
\frac{dV(\rho(t))}{dt} \leq -2\epsilon V(\rho(t)).
\label{exponential_shit}
\end{equation}
The inequality~(\ref{exponential_shit}) indicates that the transverse dynamics converges exponentially to zero during the maneuver execution, with decay rate $\epsilon$. Following established knowledge, finding an appropriate $P$ amounts to finding a positive-definite $P$ that satisfies
\begin{equation}
A(\theta)^TP+P(A(\theta)) \leq -2\epsilon P 
\label{decay_rate_shit}
\end{equation}
for all admissible values of $\theta$. The existence of such a $P$ can be obtained via a straightforward application of the circle criterion: $A(\theta) = A_0+\theta A_1$ and $A_1$ is a rank one matrix, that is $A_1= bc$, where $b$ (resp. $c$) are columns (resp. row) vectors of appropriate dimension.  Therefore, given a range of values $\left[ \theta_{\rm min}\;\;\theta_{\rm max}\right]$, the exponential stability of the system%~(\ref{transverse_shit}) 
is guaranteed if there exists $\eta>0$ such that
\begin{equation}
\left|c(j\omega I-\tilde{A}-\epsilon I)^{-1}b\right|<\frac{1}{\theta_r^2},
\label{nyquist_shit}
\end{equation}
where $\tilde{A} = A_0 + (\theta_{\rm max} - \theta_{\rm min})A_1$ and $\theta_r = (\theta_{\rm max} - \theta_{\rm min})/2$. Thus the circel criterion can be checked easily by computing the appropriate Nyquist plot and looking for its maximum modulus. Since $\tilde{A}$ is stable, it is always possible to find $\epsilon>0$ such that~(\ref{nyquist_shit}) is satisfied. Computing $P$ satisfying~(\ref{decay_rate_shit}) can then be done by solving the Riccati equation
\[
(\tilde{A}+\epsilon I)^TP+P(\tilde{A}+\epsilon I)+ Pbb^TP\theta_r^2+c^Tc = 0.
\]
The closed-loop system model below satisfies the circle criterion and therefore displays "maneuver stability".

\newpage
% \clearpage
\section*{Implementation and Validation}
\begin{figure}[h!]
\begin{center}
\includegraphics[width=\columnwidth]{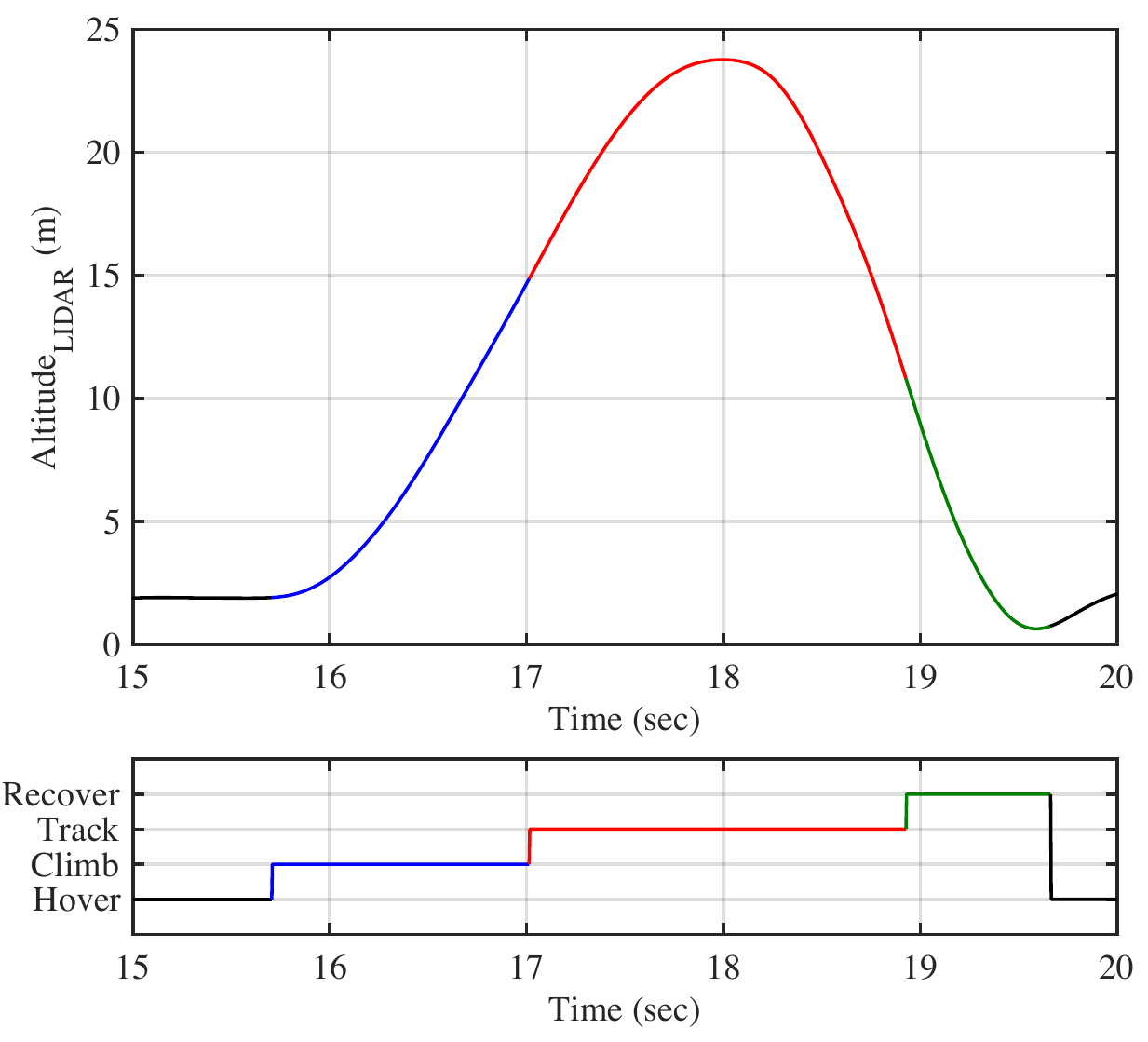}
\caption{Filtered Altitude and Automata. Different colors represent the active autonomous state during the maneuver}
\label{MartianParabola}
\end{center}
\end{figure}

On July 14th 2017, after much anticipation and tremendous efforts, the worlds first autonomous Martian gravity parabola (0.378Gs) was performed, meeting tolerances of $\pm 0.1 G's$ for a period of approximately 1.5 seconds, with a mean G of 0.3804 and a standard deviation of 0.0426. The Martian parabola performed by the vehicle is illustrated in Figure~\ref{MartianParabola}, where the LIDAR Lite data has been processes by a double sided filter. The different colors of the parabola represent the active autonomous state during the maneuver, also described at the bottom of Figure~\ref{MartianParabola} 

\begin{figure}[h!]
\begin{center}
\includegraphics[width=\columnwidth]{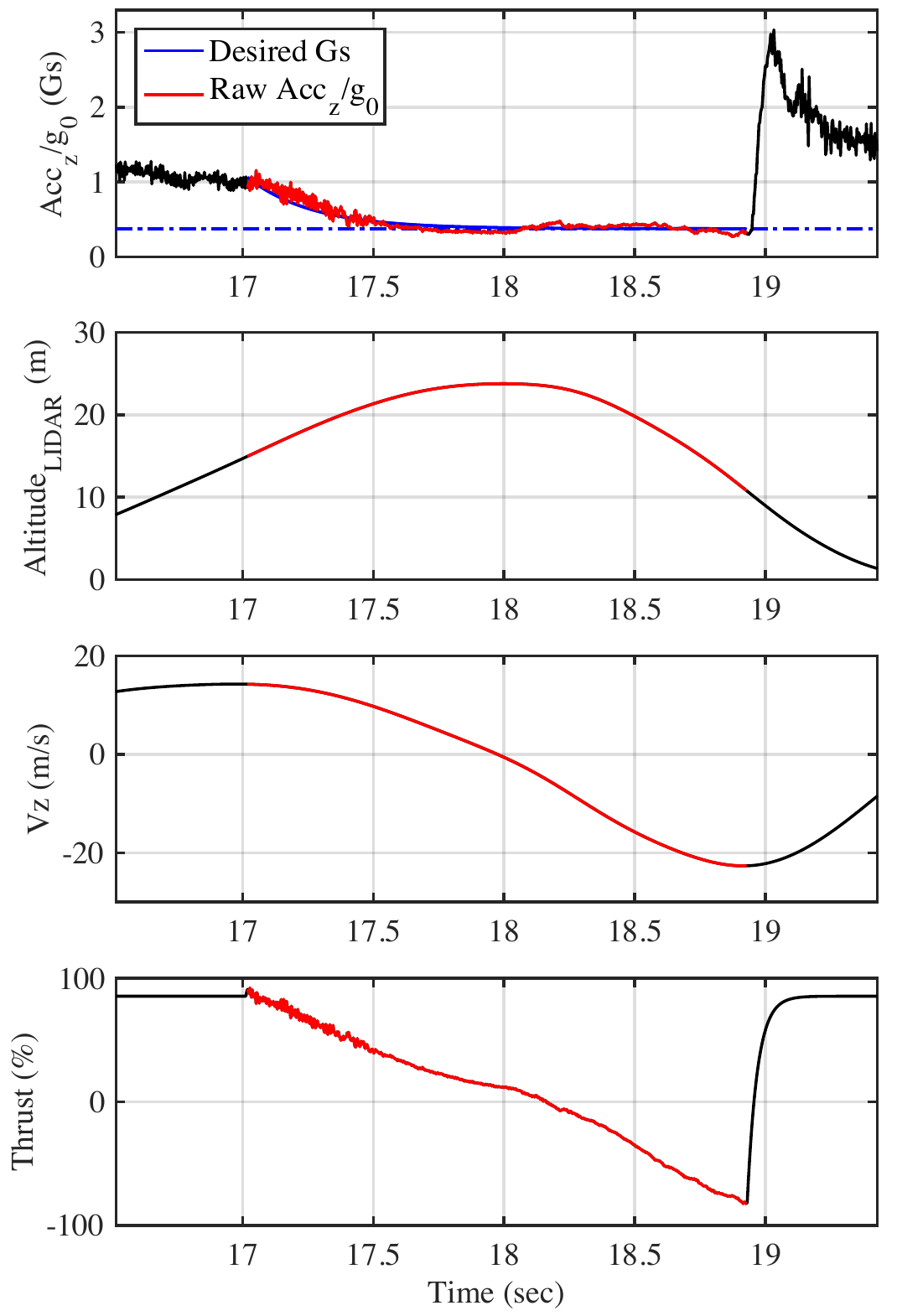}
\caption{Martian 0.378 G maneuver} %\href{https://youtu.be/MYT40zoNXuc}{(Click to see Flight Test)}}
\label{G38}
\end{center}
\end{figure}

Most importantly, Figure \ref{G38} validates the martian parabola by illustrating the transient of the raw accelerometer data, normalized by $g_0 = -9.807 m/s^2$, during the tracking maneuver shown in red. The vehicle's altitude, velocity and percent throttle during the maneuver are also illustrated. %Readers may access a video of the flight test by clicking on the label of Figure \ref{G38}. 

Although the desired acceleration was set to 0.378 G's, the reader should note that input shaping was performed by employing a first order transfer function with DC gain of one. Its purpose is to begins the maneuver by taking the current accelerometer value and gently driving it towards the desired G value based on the transfer function's time constant. The input shaping is can be is highlighted by the desired G blue line in figure \ref{G38}. %Values for the integrators of the PIRQ controller were stored during the autonomous trajectory and provided in Figure \ref{int}.

% \begin{figure}[h]
% \includegraphics[width=5cm]{3.pdf}
% \caption{First, second and third integrators from PIRQ controller during descent }
% \label{int}
% \end{figure}

While this data shows a successful implementation of the PIRQ controller, the authors acknowledge that higher fidelity acceleration tracking can be achieved by properly initializing each of the three integrators. It is also important to note that during the flight tests, strong winds were observed.  Lastly, there are unmodeled dynamics from the RPM governor controller that could be introducing instabilities along the z translation, which will have to be revised in the future.

%  \newpage
%  \clearpage

%\subsection*{Flight test}
%The model-based development described in the previous sections was employed for generating a virtual validation of the attitude and trajectory regulating control law. Once the team further confirmed the accuracy of the simulator through several flight tests, the Mathworks code generation was employed to embed the flight control software, developed for the simulator, to the Pixhawk flight control board~\cite{px4_simulink}. This hardware was chosen based on its ability to interact with the system and support the automatically generated control algorithms. The Pixhawk flight-control board features a 32bit STM32F427 Cortex M4 core with FPU and sensor technology from ST Micro-electronics. It contains a dual IMU system where an Invensense MPU 6000 supplements an ST Micro LSM303D accelerometer to provide redundancy and improve noise immunity of the power supplies. The HobbyKing Assault Reaper 500 Variable Pitch Quadrotor was selected as the experimental testbed. The test vehicle is illustrated through Fig. \ref{testbed}, shown performing an aggressive maneuver.
%
%\begin{figure}[h]
%\begin{center}
%\includegraphics[width=\columnwidth]{flight.png}
%\caption{Experimental vehicle performing aggressive maneuver}
%\label{testbed}
%\end{center}
%\end{figure}
%
%The Assault Reaper 500 is one of three variable pitch multi-rotors available in the market to date. For the experimental test, the motor and pinion gear were upgraded from stock, and the gyro was replaced by the Pixhawk flight control board.

\section*{CONCLUSION}
The objective of this paper was to present the controller design and implementation for regulating reduced-gravity maneuvers on a custom variable-pitch multirotor.  This work has validated the proposed PIRQ controller both theoretically, using the circle criterion, and experimentally on a flight test vehicle.  Future work will involve further improving the hardware and controller design to increase the accuracy of the maneuver.

% \section*{ACKNOWLEDGMENT}

% The authors wish to acknowledge themselves for being fucking awesome.

\end{document}